\documentclass[pra, article, twocolumn]{revtex4-1}

\usepackage{graphicx}
\usepackage[colorinlistoftodos]{todonotes}
\usepackage{ulem} 
\usepackage{graphicx}
\usepackage{xcolor}
\begin{document}

\title{Multimode collective scattering of light in free space by a cold atomic gas}
\author{R. Ayllon}
\affiliation{IPFN, Instituto Superior T\'{e}cnico de Lisboa, Universidade de Lisboa, Lisboa, Portugal}
\author{A. T. Gisbert}
\affiliation{Dipartimento di Fisica "Aldo Pontremoli", Universit\`{a} degli Studi di Milano, Via Celoria 16, I-20133 Milano, Italy}
\author{J.T. Mendon\c{c}a}
\affiliation{IPFN, Instituto Superior T\'{e}cnico de Lisboa, Universidade de Lisboa, Lisboa, Portugal}
\author{N. Piovella}
\affiliation{Dipartimento di Fisica "Aldo Pontremoli", Universit\`{a} degli Studi di Milano, Via Celoria 16, I-20133 Milano, Italy}
\author{G.R.M. Robb}
\affiliation{SUPA and Department of Physics, University of Strathclyde,Glasgow G4 0NG, Scotland, UK}

\begin{abstract}
We have studied collective recoil lasing by a cold atomic gas, scattering photons from an incident laser into many radiation modes in free space. The model consists of a system of classical equations for the atomic motion of $N$ atoms, where the radiation field has been adiabatically eliminated. We performed numerical simulations using a molecular dynamics code, Pretty Efficient Parallel Coulomb Solver or PEPC, to track the trajectories of the atoms. These simulations show the formation of an atomic density grating and collective enhancement of scattered light, both of which are sensitive to the shape and orientation of the atomic cloud. In the case of an initially circular cloud, the dynamical evolution of the cloud shape plays an important role in the development of the density grating and collective scattering. 
The ability to use efficient molecular dynamics codes will be a useful tool for the study of the multimode interaction between light and cold gases.
\end{abstract}

\date{1 June 2019}
\maketitle

\section{Introduction : Scattering of Light by Atoms}
One of the most basic light-atom interactions is Rayleigh scattering. When an ensemble of $N$ randomly distributed, stationary atoms is weakly illuminated by a laser, the atoms scatter independently and the resultant scattered field intensity varies as $\sim N$.
For an ensemble of cold atoms which are free to move, the picture can change drastically due to collective behaviour arising from the optical forces produced during scattering.  Each atom is affected by the optical field scattered by the other atoms.
Most studies of collective behaviors involving cold and ultracold atoms coupled to  light have involved optical cavities \cite{review}, but similar phenomena have also been observed or predicted involving single feedback mirrors, optical fibres and simply scattering into vacuum. These collective behaviors are at the origin of various self-organization phenomena, e.g.  collective cooling  \cite{Ritsch,Vuletic2003,Vuletic2017}, symmetry breaking and pattern formation \cite{Esslinger,Tesio,Labeyrie,Robb,Zhang,Greenberg,Schmittberger,Greiser,Mendonca}.

Superradiant light scattering was first demonstrated using a cigar-shaped Bose-Einstein Condensate (BEC) \cite{Inouye} and later using a cold, thermal gas \cite{Yoshikawa}. In \cite{Inouye}, superradiantly scattered light was observed to propagate along the major axis of the atomic cloud, simultaneous with the development of a matter-wave/density grating in the cloud. Some features of this phenomenon have been described by single-mode/mean field models similar to that of the Collective Atomic Recoil Laser (CARL) \cite{Moore,Bonifacio1,Bonifacio2,Bonifacio4,Piovella2001,You,Zobay1,Zobay2,Chen,Slama2007,Slama2007bis,Bux2013}. These mean-field models are appropriate in certain specific cases where there is a well-defined propagation axis and consequently, to a good approximation, a single spatial mode, e.g., in a single-mode cavity or in a highly elongated sample where the major axis of the sample defines an `end-fire mode' which dominates the direction of emission. In general, however, for arbitrary shapes of atomic ensembles, many spatial modes are involved simultaneously in the collective scattering process.

The computational effort required to model large systems of atoms in 2D and 3D geometries is significant. Large efficient publicly accessible `molecular dynamics' (MD) codes, which solve dynamical equations of motion for large collections of particles under the action of various forces (gravitational, electrostatic, van der Waals), have become an essential tool in many areas of science, e.g., plasma physics, astrophysics \& computational chemistry. Despite the latter fact, to date, MD codes have not been used in the study of light interacting with cold atomic gases.

In this work we have simulated collective light scattering from a gas of cold atoms in 2D and 3D, using a model which describes the positions and velocities of the atoms. The model has been derived from a multimode theory, where the vacuum radiation modes are adiabatically eliminated. The result is a set of coupled $N$ atoms where each atom is subjected to the radiation force exerted by all the other atoms present in the cloud. The form of the equations in this model makes them suitable for implementation in MD codes, which offers the possibility of efficient simulation of multimode scattering involving very large numbers of atoms by exploiting methods developed for simulating $N$-body systems involving long-range interactions e.g., Barnes-Hut methods \cite{Barnes-Hut}. We use a public MD code, PEPC \cite{PEPC}, to demonstrate that the collective scattering process described by our model has similar characteristics to those observed in \cite{Inouye}, i.e., observation of a density grating, which is responsible for collective enhancement of scattered light intensity.
Whereas for ultracold atoms the grating is observed in momentum space \cite{Inouye}, with spacing $\hbar\mathbf{q}=\hbar(\mathbf{k}_0-\mathbf{k})$ $\mbox{---}$where $\hbar\mathbf{k}_0$ and $\hbar\mathbf{k}$ are the momentum of the incident and scattered photon$\mbox{---}$, here, in contrast, the grating is observed in real space, with atoms grouping periodically at distances which are multiples of $2\pi/q$.
The model employed to depict the evolution of the cloud is presented in Section \ref{model}, along with its implementation in the MD algorithm. It is possible to see that by using particular atomic cloud shapes and orientations, different density grating shapes and scattered light directions are achieved. These results, for a 2D cloud and for a specific 3D geometry, are presented in Section \ref{results:2d} and \ref{results:3d}, respectively.

\section{Model of Collective Scattering}
\label{model}

We consider a collection of $N$ two-level atoms driven by a laser field with frequency $\omega_0=ck_0$, propagating along the $z$-axis with wave number $\mathbf{k}_0=k_0\hat\mathbf{z}$ and Rabi frequency $\Omega_0=dE_0/\hbar$, where $E_0$ is the electric field and $d$ is the atomic dipole. The laser field is far detuned from the atomic frequency $\omega_a$, with $\Delta_0=\omega_0-\omega_a\gg\Gamma$, being $\Gamma=d^2k_0^3/2\pi\epsilon_0\hbar$ the atomic linewidth. In the far-detuned limit and for a dilute gas, absorption and multiple scattering can be neglected. In this limit, the incident light in the mode $\mathbf{k}_0$ is scattered into the vacuum mode $\mathbf{k}$. The scattered optical field in the mode $\mathbf{k}$ interferes with the incident mode $\mathbf{k}_0$ to create a dipole force proportional to the photon momentum transfer $\hbar(\mathbf{k}_0-\mathbf{k})$. When summed over the different  vacuum modes, the resulting equations for atomic positions $\mathbf{r}_j$ and momenta $\mathbf{p}_j$ are (see Appendix \ref{appendix}):
\begin{widetext}
\begin{eqnarray}
\dot \mathbf{r}_j&=&\frac{\mathbf{p}_j}{M},\label{rj}\\
\dot \mathbf{p}_j
&=&\Gamma\hbar k_0\left(\frac{\Omega_0}{2\Delta_0}\right)^2\sum_{m\neq j}\left\{
(\hat\mathbf{z}-\hat\mathbf{r}_{jm})
\frac{\sin[k_0(r_{jm}-z_{jm})]}{k_0r_{jm}}-\hat\mathbf{r}_{jm}\frac{\cos[k_0(r_{jm}-z_{jm})]}{(k_0r_{jm})^2},
\right\},
\label{pj}
\end{eqnarray}
\end{widetext}
where $M$ is the atomic mass, $\mathbf{r}_{jm}=\mathbf{r}_j-\mathbf{r}_m$ and $\hat\mathbf{r}_{jm}=\mathbf{r}_{jm}/r_{jm}$. Each atom, labelled $j$, is coupled to all the other $m$-atoms (where $m \neq j$) by an oscillating force with components along the direction $\hat\mathbf{z}$ of incident field  and the direction $\hat\mathbf{r}_{jm}$ toward the other atoms. Furthermore, the force has a finite range, consisting of terms which decrease with distance between the atoms as $1/r_{jm}$ or $1/r_{jm}^2$. 

The intensity of scattered light in the direction ${\mathbf k}$ is
\begin{eqnarray} \label{intensity}
 I_s(\mathbf{k})&=&I_1N^2 |M(\mathbf{k},t)|^2,
 \end{eqnarray}
where $I_1=(\hbar\omega_0\Gamma/8\pi r^2)(\Omega_0/2\Delta_0)^2$ is the single-atom Rayleigh scattering intensity and
\begin{eqnarray} \label{bunching_factor}
M(\mathbf{k},t)&=&\frac{1}{N}\sum_{j=1}^N e^{i(\mathbf{k}_0-\mathbf{k})\cdot\mathbf{r}_j(t)}
 \end{eqnarray}
is the 'optical magnetization', or 'bunching factor'. It describes the strength of the density grating formed by the moving atoms; ranging from zero, when the atomic positions are uniformly distributed, to unity, when the atoms are periodically packed into a length less than  $2\pi/|\mathbf{k}_0-\mathbf{k}|$. These equations generalize the Collective Atomic Recoil Laser (CARL) model, obtained for atoms interacting with a single mode in an optical ring cavity \cite{Bonifacio1}, to many modes in vacuum. Here the incident photons are scattered in the 3D vacuum, and superradiant scattering occurs along certain directions determined by the atomic spatial distribution. In particular, for an elongated atomic distribution along the $z$-axis of the incident field, collective scattering occurs along the backward direction $\mathbf{k}=-\mathbf{k}_0$.

The present model assumes a scalar radiation field, disregarding polarization effects. This approximation can result in an inaccurate description of the scattered light and/or the radiation force among the atoms, particularly in the case of a 3D atomic distribution. However, a full derivation of the vectorial light model (not presented here) shows that the scalar light model describes correctly the long-range contribution$\mbox{---}$i.e., the first term of the force in the right-hand side term of Eq.(\ref{pj})$\mbox{---}$, for a pump linearly polarized in a direction perpendicular to the scattering plane. Differences between the vectorial and scalar light models arise only in the short-range terms of the radiation force, which are less important in the collective recoil regime considered here. A detailed study of collective scattering using the vectorial light model will be the subject of a future publication.

\subsection{Simulation Algorithm}
Due to the form of Eqs.(\ref{rj}) and (\ref{pj}), it is possible to simulate collective light scattering using a molecular dynamics (MD) code. We used the Pretty Efficient Parallel Coulomb Solver (PEPC) [8], which is commonly used for simulating N-body systems where the forces involved are described by an inverse-square law, e.g., Coulomb forces in plasmas and gravitational forces. In order to model collective scattering of light by atoms, we implemented Eq. (\ref{pj}) as the force acting on each atom, and observed the trajectories of the particles. Since the equations only depend on the positions of the particles, the force for each iteration was calculated using the position Verlet algorithm, which updates the position of each atom according to 
\begin{equation}
\label{eqn:Verlet}
\mathbf{r}_{n+1} = 2 \mathbf{r}_{n}- \mathbf{r}_{n-1} + \mathbf{a}_n \Delta t^2,
\end{equation}
where $\mathbf{a}_n$ is the acceleration at time step $n$.
The Verlet integrator provides good numerical stability, as well as other properties that are important in physical systems such as time reversibility and preservation of the symplectic form in phase space.
The form of the model equations shows a singularity when the particles are close to each other. This becomes an important issue during the simulation, since it results in strong forces appearing abruptly, causing the particles to be ejected from the cloud, i.e., two atoms repel one another violently when they get too close to each other. We solved this problem using the idea of Plummer \cite{Plummer}, which is used in gravitational force simulations, and involves making the replacement: 
\begin{equation}
\label{eqn:Plummer}
r_{jm} \rightarrow \sqrt{r_{jm}^2 + \epsilon^2}~,
\end{equation}
where $\epsilon$ is a small parameter introduced in order to avoid singularities in the equations. This parameter does not change the general behaviour of the system when the particles are well separated. It just allows the particles to pass each other as if they were experiencing an elastic collision characterized by the parameter ($\epsilon$), which in some sense acts as a numerical scattering length. This collision could be interpreted as a repulsion generated due to van der Waals forces between a pair of atoms.

The equations have been scaled in order to work with dimensionless variables. Positions have been scaled like  $\mathbf{r}' = k_0\mathbf{r} $; the momentum variable as $\mathbf{p}' = \mathbf{p} p^{-1}_0$, where $p_0 = \hbar k_0$ is the momentum of a single photon; and the time variable like $t' = \omega_r t$, where $\omega_r = \hbar k^2_0/2m$ is the recoil frequency. Introducing these variables into the equation of motion (\ref{rj}) and (\ref{pj}) we obtain equations
\begin{widetext}
\begin{eqnarray}
\dot \mathbf{r}'_j&=& 2 \mathbf{p}'_j~,\label{rj_adimention}\\
\dot \mathbf{p}'_j&=&
A\sum_{m\neq j}\left\{
(\hat\mathbf{z}-\hat\mathbf{r}_{jm})
\frac{\sin[r'_{jm}-z'_{jm}]}{r'_{jm}}-\hat\mathbf{r}_{jm}\frac{\cos[r'_{jm}-z'_{jm}]}{(r'_{jm})^2}
\right\},
\label{pj_adimention}
\end{eqnarray}
\end{widetext}
with
\begin{equation}
A = \frac{\Gamma}{\omega_r} \left(\frac{\Omega_0}{2\Delta_0} \right)^{2},
\end{equation}
and $r'_{jm} \rightarrow \sqrt{r'_{jm} + \epsilon'^2}$, where the singularity-avoiding parameter becomes $\epsilon' = k_0\epsilon$. 

The value of the singularity-avoiding parameter used in our simulations was $\epsilon' = 10^{-2}$. This implies that the atoms in our simulation have an effective scattering length $\sim 10^{-2} \lambda_0$, where $\lambda_0=2\pi/k_0$ is the laser wavelength.
Regarding other important variables, we have used $\omega_r \approx 10^4 s^{-1}$, as the recoil frequency, $\Gamma \approx 10^7 s^{-1}$, for the atomic decay rate, and we have selected $A = 1.0$ for simplicity. By choosing these values, we roughly achieve that $\Delta_0 \approx 15 \Omega_0$, hence fulfilling the necessary conditions of the model. For both simulations in 2D, we have adopted a time step, $\delta t' = 0.15\times 10^{-3}$, with $2000$ steps, which makes a total simulation time of $t = 0.3\omega^{-1}_{r}$. Instead, for the simulations in 3D, the selected step is $\delta t' = 0.25\times 10^{-3}$ with $7000$ steps, which in turn corresponds to a total time of $t = 1.75\omega^{-1}_{r}$.

\begin{figure}[!hb]
\centering
\includegraphics[width=\columnwidth]{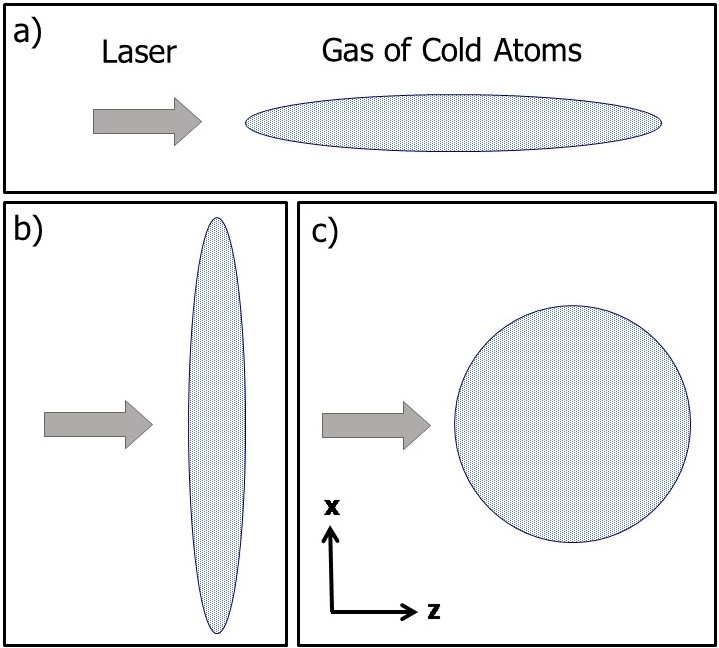}
\caption{Schema of the three different configurations used in our simulations. a) Elliptical/cigar-shaped gas of atoms with a major axis directed parallel to the propagation direction of the laser. b) The elliptical gas with major axis orientated perpendicular to the propagation direction of the laser. c) A circular-shaped atomic gas.}
\label{fig1}
\end{figure}

\section{Results}

\subsection{Simulations of the scattering from a 2D atomic cloud}
\label{results:2d}
In this section we restrict ourselves to a simplified configuration where the atomic distribution is two-dimensional, consisting of two geometries: an ellipse (sections \ref{results:2d_parallel} and \ref{results:2d_perp}) and a circle (section \ref{results:2d_circ}), with both distributions being contained in the $(x,z)$ plane. It is well known from experimental studies of superradiance and superfluorescence, both in excited atomic systems of effectively stationary atoms \cite{atomicSR} and in BECs \cite{Inouye,Yoshikawa,Schneble}, that the geometry of the atomic cloud/sample can have a significant effect on the spatial distribution of the emitted field. We will demonstrate that the spatial distribution of both the scattered radiation and the associated atomic density distribution, which is produced during collective scattering of light, are also strongly affected by the geometry of the atomic cloud. 

\subsubsection{Pump propagation parallel to the major axis of an elliptical cloud : Backscattering and 1D grating formation}
\label{results:2d_parallel}
\begin{figure}[!ht]
\begin{center}
\includegraphics[width=\columnwidth]{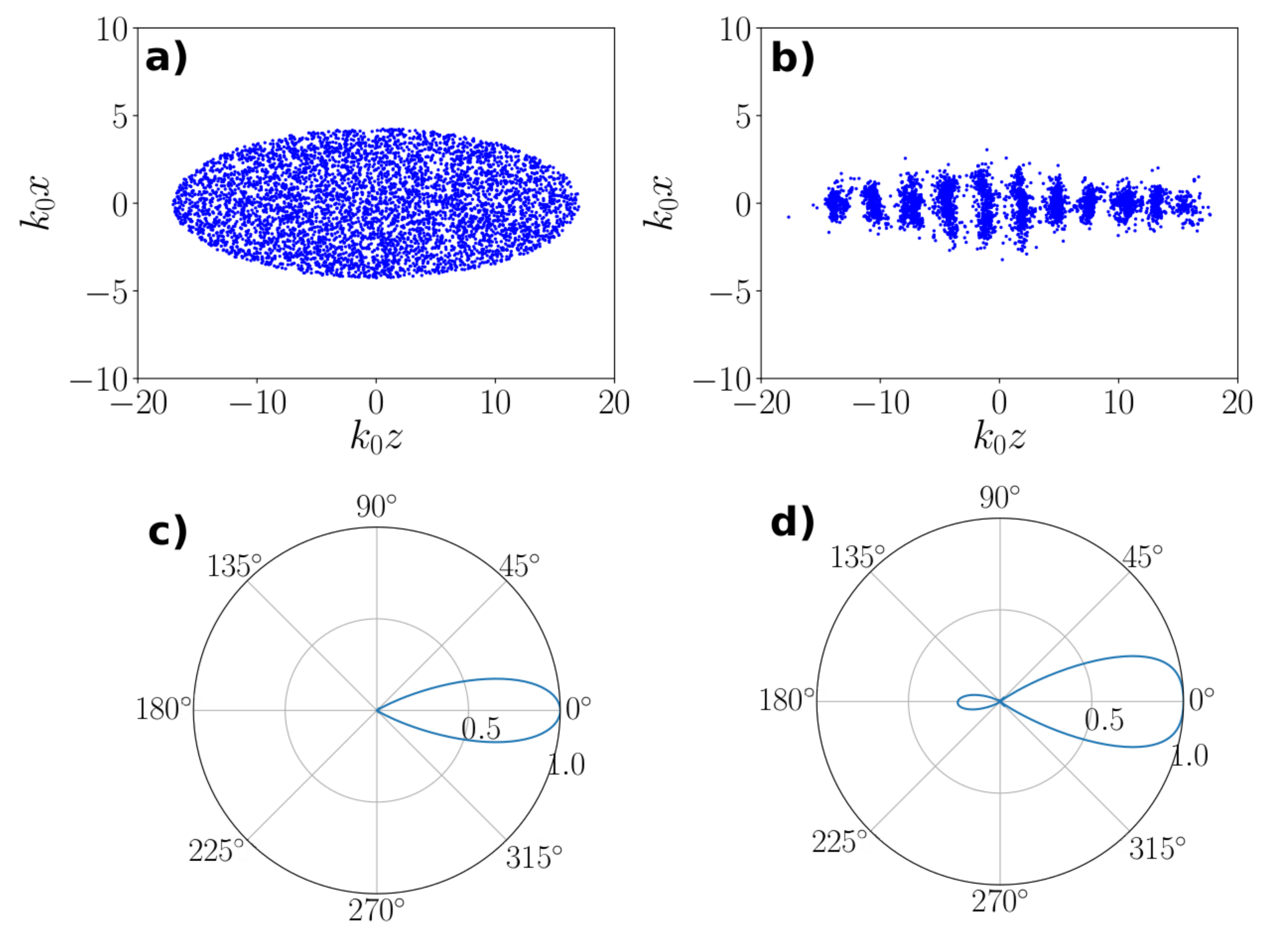}
\end{center}
\caption{Simulation of collective scattering of a pump laser propagating parallel to the major axis of an elliptical 2D atomic cloud: a) Initial atomic density distribution showing $N \approx 5000$ particles distributed randomly. b) Density grating formation due to collective scattering at $t = 0.135\omega^{-1}_{r}$. The corresponding bunching factors, $|M(\mathbf{k},t)|$ are shown in (c) at $t=0$ and in (d) at $t = 0.135\omega^{-1}_{r}$.}
\label{fig2}
\end{figure}

The first case we examine is that of an elliptical atomic cloud illuminated by an optical pump field whose propagation direction is parallel to the major axis of the cloud as shown schematically in Fig.~\ref{fig1}(a).  Fig.~\ref{fig2}(a) shows the initial, random, distribution of atoms in the atomic cloud. As a consequence of the optical forces arising from Rayleigh scattering, this initially random spatial distribution of atoms develops a strong periodic modulation along the $z$-direction, with a spatial period $\approx \lambda/2$, as shown in Fig.~\ref{fig2}(b). Consequently, the atomic cloud undergoes the spontaneous formation of a 1D density grating, analogous to the ones occurring in CARL or a free electron laser (FEL). Observing Fig.~\ref{fig2}(d) we conclude that the 1D grating forms because light is predominantly backscattered, due to the geometry of the atomic cloud, which leads to scattering along the cloud's major axis in both the $\pm z$ directions. Light which is forward scattered in the $+ z$ direction will not produce an optical force on an atom, as there is no change of the photon momentum during scattering. We remember that we have neglected the effect of the scattering force, in the limit of large detuning $\Delta_0\gg\Gamma$ (Appendix), which eventually pushes the atoms in the direction of the pump \cite{Bienaime2010}. Conversely, light backscattered along the $- z$ direction produces an optical force on an atom, as the optical field propagation direction and consequently momentum changes during the scattering process. This change in momentum of the optical field is taken up by an atom, moving it and  modifying the atomic density. The backscattered light interferes with the pump field to form a 1D optical potential with a spatial period of $\approx \pi/k_0$, that has an amplitude and a position which evolve dynamically, and consistently, with the developing atomic density modulation.

The forward lobe of the scattered intensity in Fig.~\ref{fig2}(c) is the result of the diffraction by the atoms in the initial distribution. For a uniform ellipse with semiaxis $R_x$ and $R_z$, the bunching factor $|M(\theta,\phi)|$ is
\begin{equation}
|M(\theta,\phi)|=\frac{2J_1\left[k_0\sqrt{R_x^2\sin^2\theta\cos^2\phi+R_z^2(1-\cos\theta)^2}\right]}
    {k_0\sqrt{R_x^2\sin^2\theta\cos^2\phi+R_z^2(1-\cos\theta)^2}},
\end{equation}
where we assumed $\mathbf{k}=k_0(\sin\theta\cos\phi,\sin\theta\sin\phi,\cos\theta)$, $\mathbf{k}_0=k_0\hat\mathbf{z}$ and $J_1(x)$ is the first-order Bessel function. The majority of the emission is within the diffraction angle $\Delta\theta \sim 1/(k_0R_x)$.

\subsubsection{Pump propagation perpendicular to the major axis of an elliptical cloud : Off-axis scattering}
\label{results:2d_perp}
\begin{figure}[!ht]
\centering
\includegraphics[width=\columnwidth]{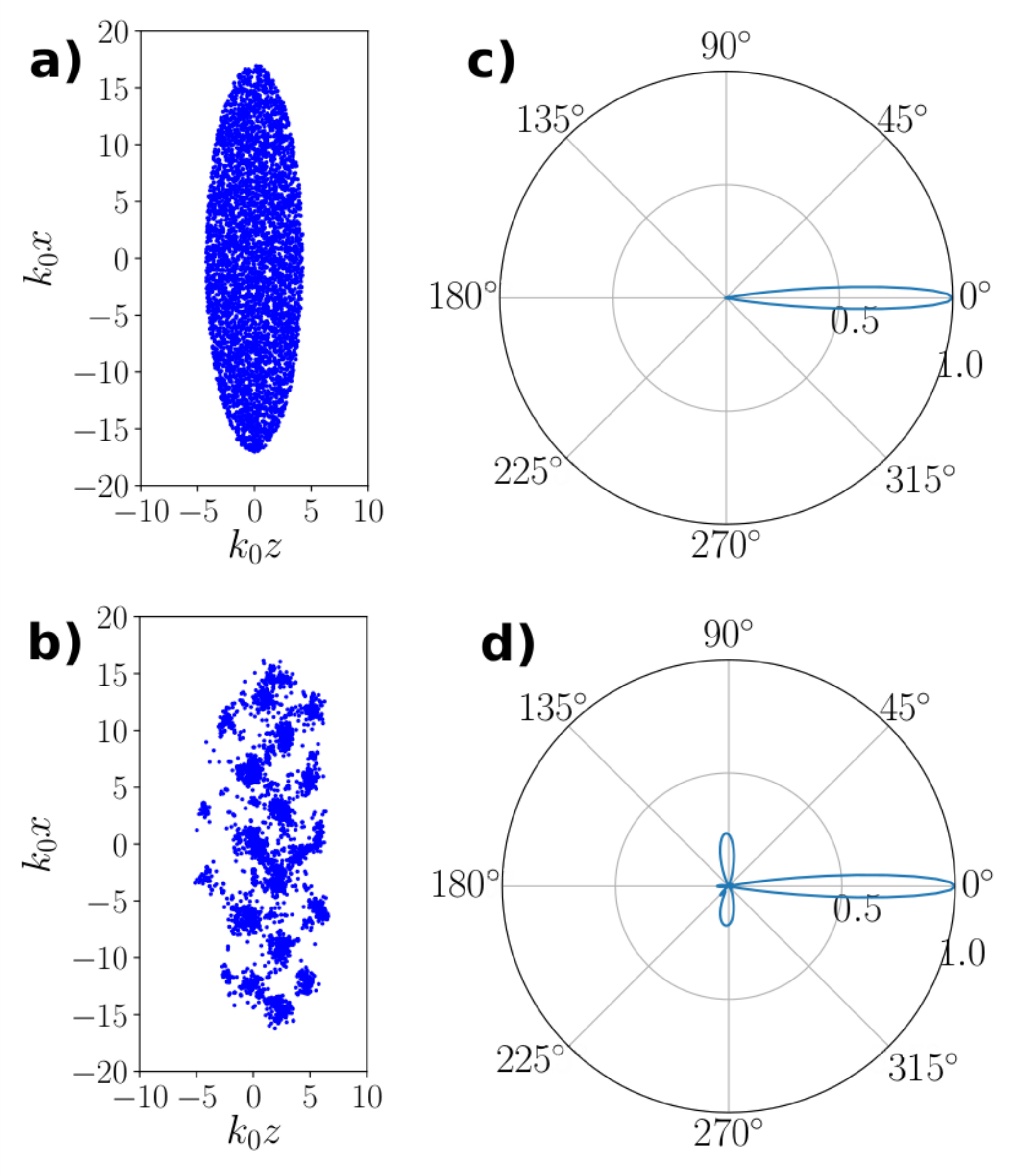}
\caption{Simulation of collective scattering of a pump laser propagating perpendicular to the major axis of a 2D elliptical atomic cloud: a) Initial atomic density distribution showing $N \approx 5000$ particles distributed randomly. b) Density grating formation due to collective scattering at $t = 0.159\omega^{-1}_{r}$.  The corresponding bunching factors, $|M(\mathbf{k},t)|$, are shown in (c) at $t=0$ and in (d)  at $t = 0.159\omega^{-1}_{r}$.} 
\label{fig3}
\end{figure}

We now consider the case where the optical pump field propagates perpendicular to the major axis of the elliptical atomic cloud, as shown schematically in Fig.~\ref{fig1}(b). The initial random distribution of atoms in the atomic cloud is shown in Fig.~\ref{fig3}(a). In this case, the initially random distribution of atoms again develops a strong periodic modulation and forms a density grating, but in contrast to the previous case of section~\ref{results:2d_parallel}, this grating is now no longer restricted to the $z$-axis, but is a 2D structure in the $(x,z)$ plane. Figs.~\ref{fig3}(c,d) show that the 2D grating forms because the geometry of the atomic cloud, which leads to significant scattering perpendicular to the pump propagation direction, along the major axis of the atomic cloud in both the $\pm x$ directions. Scattering of light along the $\pm x$ directions will produce an optical force on an atom directed at approximately $\mp 45^\circ$ to the $z$-axis. This can be understood using a photon picture of a scattering event which involves an incident photon with momentum $(\hbar k_0)\hat{\mathbf{z}}$ and results in a scattered photon of momentum $(\hbar k_0) \hat{\mathbf{x}}$. This results in a net momentum change of the atom of $\hbar k_0 (\hat{\mathbf{z}} \mp \hat{\mathbf{x}})$, i.e., directed at approximately $\mp 45^\circ$ to the $z$-axis, depending whether the photon is emitted upward or downward, respectively. This scattered light interferes with the pump field to form a dynamically evolving, 2D optical lattice potential \cite{Piovella2001}. An atomic density distribution similar to that shown in Fig.~\ref{fig3}(b) was observed by Inouye et al. \cite{Inouye} for the case of an elongated, elliptical Bose-Einstein condensate (BEC), illuminated by a pump beam propagating perpendicular to its major axis. Whereas in the experiment of Ref.\cite{Inouye} the grating is observed in momentum space, after the interaction with the pump laser, here the grating is observed in  real space. 

\subsubsection{Scattering from a circular atomic distribution}
\label{results:2d_circ}
\begin{figure}[!ht]
\centering
\includegraphics[width=\columnwidth]{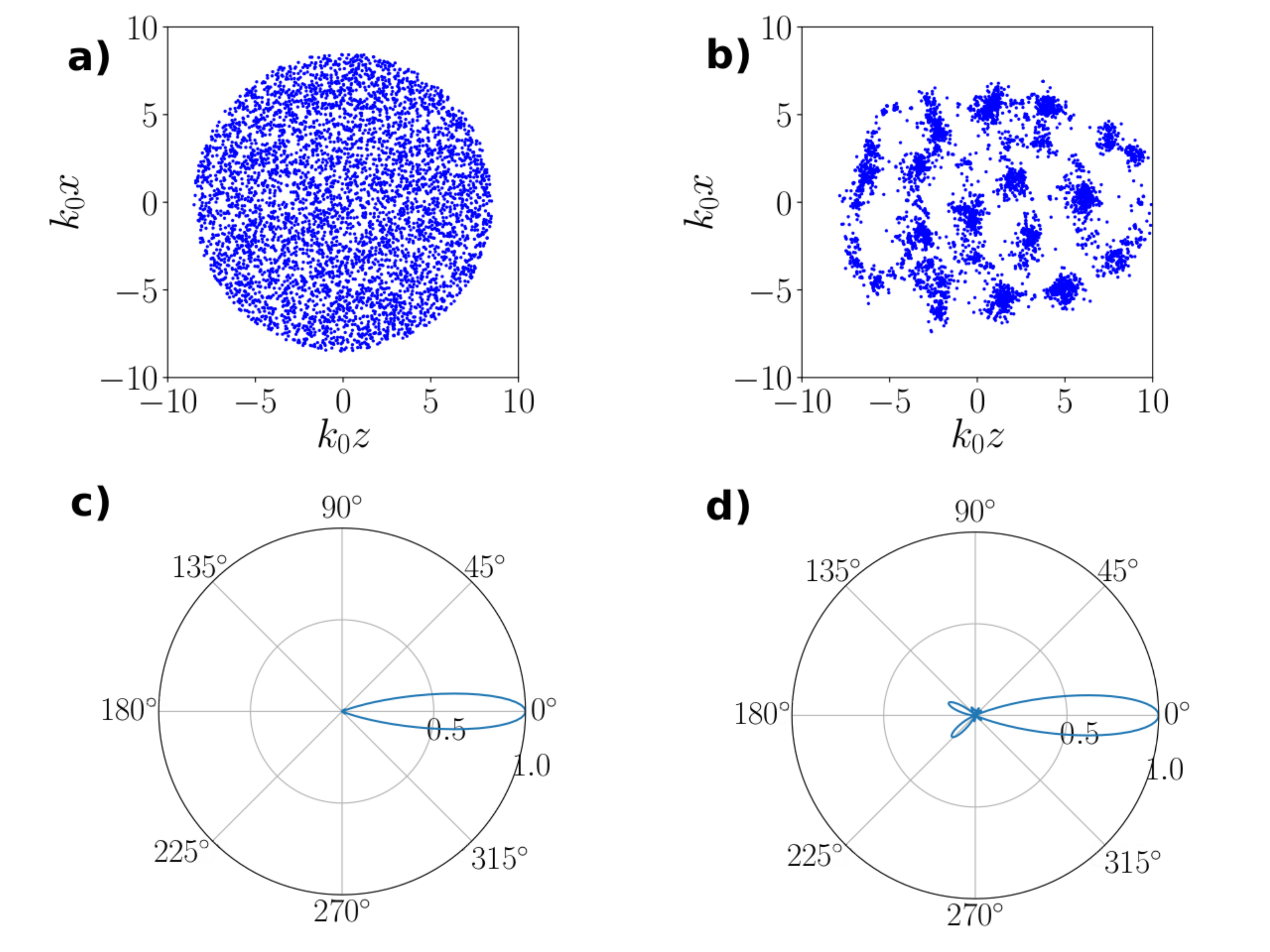}
\caption{Simulation of collective light scattering from a 2D circular atomic cloud : a) Initial atomic distribution showing $N \approx 5000$ particles distributed randomly. b) Atomic distribution at $t = 0.22\omega^{-1}_{r}$.  The corresponding bunching factors, $|M(\mathbf{k},t)|$, are shown in (c) at $t=0$ and in (d)  at $t = 0.22\omega^{-1}_{r}$.} 
\label{fig4}
\end{figure}

We now consider the light scattering from the  circular 2D simulation shown in Fig.~\ref{fig1}(c). Since now there is not any preferred scattering direction, we would expect to observe no density grating in this case. Instead, we can still see the formation of a 2D grating due to a periodic modulation. Observing the polar plot that represents the bunching parameter for this configuration, Figs.~\ref{fig4}(c,d), we can see that, at a certain time, the cloud scatters light in two directions, at approximately $\pm 45^\circ$ from the backward direction. This can be interpreted taking into consideration the deformation of the initially round distribution. It can be observed in Fig.~\ref{fig4}(b) that the atoms close to the $z$-axis and on the right edge of the initial distribution are pushed forward by the pump laser, making the atomic cloud form an 'egg-like' shape. Since scattered light is preferentially amplified along the longest propagation path in the cloud, this path results in being along the edges of the egg-like shape formed after an initial transient time. If we look at the deformed shape in Fig.~\ref{fig4}(b) as if it was a triangle with two equal angles (located at the negative plane of z-axis) and a third one (placed on the positive z-axis) that would identify the angle between the two scattered light directions.
Naming $\theta$ the angle of the scattered light direction with respect to the $z$ axis, we still interpret a scattering event using a photon picture:  the incident photon with momentum  $\mathbf{q}_{in}=\hbar k_0 \hat{\mathbf{z}}$ is scattered in the directions $\pm\theta$ as a photon of momentum 
$\mathbf{q}_{\pm}=\hbar k_0 [\hat{\mathbf{z}} \cos\theta \pm \hat{\mathbf{x}}\sin\theta]$, respectively.
The atomic recoil momentum is 
\begin{equation}
\Delta \mathbf{p}=\mathbf{q}_{in}-\mathbf{q}_{\pm}=\hbar k_0 [\hat{\mathbf{z}} (1 -\cos\theta) \mp \hat{\mathbf{x}}\sin\theta]\label{dp},
\end{equation}
with an angle $\phi$ respect to the $\mathbf{z}$ axis given by
\begin{equation}
\tan\phi=\mp \frac{\sin\theta}{1-\cos\theta}\label{angle}
\end{equation}
The previous cases of horizontal and vertical ellipses, shown in fig.\ref{fig2} and \ref{fig3}, correspond to $\theta=\pi$ and $\theta=\pi/2$, respectively.
For the case of circular distribution, we estimated from fig.\ref{fig4}(d) the scattering angle to be $\theta\approx 135^\circ$. Using this value in Eq.(\ref{angle}) we obtained two crossed lattices, oriented respectively at $\phi=\mp 22.5^\circ$ with respect to the $z$ axis, in qualitative agreement with Fig.~\ref{fig4}(b)
The shape deformation of the atomic distribution observed here is similar to the electrostrictive effect described in \cite{Kurizki} for a BEC illuminated by laser light. We postpone the study of this rather surprising effect to a more extended 2D and 3D investigation, which will take into account also the vectorial character of the scattered light.

\subsection{3D simulation of scattering}
\label{results:3d}

\begin{figure*}[!htp]
\centering
\includegraphics[width=\textwidth]{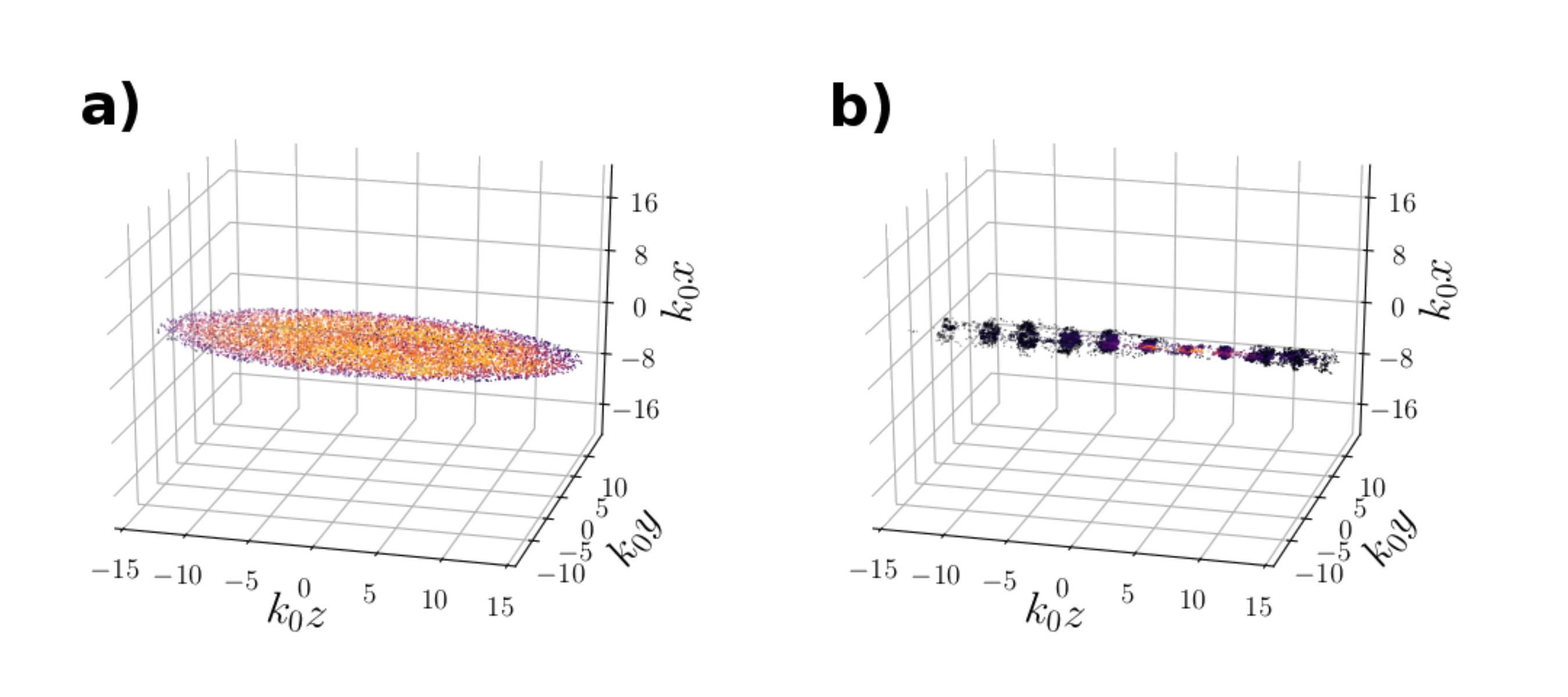}
\caption{Numerical simulations in 3D: (a) Initial disposition of particles in a cloud of particles. (b) 1D grating formation in the case of laser propagation parallel to the major axis of the cloud of atoms at $t=0.21\omega^{-1}_{r}$. In the simulation we have used $N\approx10000$ particles distributed randomly in space.}
\label{fig5}
\end{figure*}

In this section, we relax the assumption of a 2D distribution of atoms and consider a full 3D case. The computational effort required to model large systems of atoms in 3D is substantially greater than in 2D, so the efficiency of the computational methods used becomes more significant. Equations (7-8) are explicit equations whose solution does not require inversion of large matrices, nor the use of a mesh, which is attractive from the viewpoint of run-time of numerical simulations,  In addition, use of a code like PEPC to solve eq. (7-8)  offers the potential for improved scalability to large 3D simulations involving extremely large numbers of over a "brute-force" solution of eq. (7-8). This is due to the fact that PEPC is designed to use tree-algorithms (e.g. [18]) originally designed for astrophysical $N$-body simulations, which reduce the computational effort or run-time associated with the calculations from O($N^2$) to O($N \log(N)$). As an illustrative example we study a 3D atomic sample, analogous to the system configuration considered in section~\ref{results:2d_parallel}, with the pump propagation parallel to the major axis of the cloud with a cigar-shaped distribution$\mbox{---}$see Fig.~\ref{fig5}(a))$\mbox{---}$; again, we have the atoms initially randomly distributed within the cloud. After a time $t=0.21\omega^{-1}_{r}$, we observe the formation of a longitudinal density grating along the $z$-axis, depicted in Fig.~\ref{fig5}(b)), similar to the one observe in the 2D simulation. We outline again that the scalar model of light used for the 3D simulations gives only an approximated description of the scattering, so that a full vectorial model is required for an accurate description of the scattering. However, preliminary results show that for very elongated atomic cloud and the pump propagating along the major axis of the cloud, the scalar model describes correctly the long-range term of the exact force but not its short-range terms. Since we assume a dilute system, where multiple scattering is negligible, short-range terms in the force play a minor role, and the collective recoil scattering is dominated by long-range interactions. For these reasons, the scalar model is able to reproduce the mean features of the collective atomic recoil lasing in free space.

Finally, we make some comment about the scaling laws with $N$ and the size of the atomic cloud. In our 3D simulation, the number of atoms is $N=10^4$ and the semi-axis of the ellipsoidal are $k_0R_x=k_0R_y\sim 5$ and $k_0R_z\sim 15$, corresponding to a volume of $V\sim 6 \lambda_0^3$ which, for $\lambda_0=780$nm as for the Rb atoms, conforms a rather unrealistic density of $n\sim 10^{15}$ atoms/cm$^3$ and a resonant optical thickness of $b\sim N/(k_0^2R_xR_y)\sim 400$, which is large but not unreachable. Hence, it is important to know how the superradiant scattering rate scales with $N$ and the atomic system size. It results from a single-mode theory \cite{Piovella2001} that the superradiant scattering rate is $\Gamma_{SR}\sim (\Omega_0/\Delta_0)\sqrt{\Gamma\omega_r N/(k_0R_x)^2}$, i.e., it scales with the square root of the optical thickness. Hence, a realistic atomic cloud with $N\sim 10^6$ and transverse size $k_0R_x\sim 50$ would have the same optical thickness and hence the same superradiant rate of the simulation shown in Fig.~\ref{fig5}. For $\Omega_0/\Delta_0=1/15$, $N=10^6$, $k_0R_x=50$, $\omega_r\sim 10^4$ rad/s and $\Gamma=(2\pi)\, 6$MHz, then $\Gamma_{SR}\sim 10^6$ 1/s, which is much more than the two-photon recoil $4\omega_r$, and so satisfying the condition for the classical regime of superradiant scattering \cite{Gatelli}. The initial velocity spread is negligible if $2k\sigma_v\ll\Gamma_{SR}$ or equivalently $T_0\ll \hbar\Gamma_{SR}^2/(8k_B \omega_r)$, where $T_0=M\sigma_v^2/k_B$ is the initial temperature and $k_B$ is the Boltzmann's constant. For Rb atoms and $\Gamma_{SR}=10^6$ 1/s, the initial temperature must be much less than $100\,\mu$K.

\section{Conclusions}
We have presented a model which describes collective scattering of light in 2D/3D due to a gas of cold atoms in vacuum that depends only on the positions of the atoms, making it suitable for implementation using a MD simulation code. Using the public MD code, PECP, we were able to follow the trajectories of the atoms and calculate the spatial and temporal evolution of the intensity of the scattered light. 
The 2D simulations show that the evolution of collective scattering by an elliptical atomic cloud is sensitive to the orientation of the cloud relative to the pump field propagation direction. When the major axis of the cloud is aligned parallel to the pump propagation direction, the simulation showed formation of a 1D grating in the density of the atoms, analogous to that occurring in CARL or a free electron laser (FEL), which enhances the backscattered light.
In contrast, when the major axis of the cloud is oriented perpendicular to the pump propagation direction, a 2D pattern formation, similar to that observed in superradiant scattering experiments of \cite{Inouye}, was observed; in both cases, the collectively scattered radiation propagates predominantly along the major axis of the cloud. In the intermediate case of a circular cloud, it was demonstrated  that the force produced by the collective scattering process is electrostrictive in nature, leading to elongation of the cloud along the pump propagation direction, simultaneous with the development of a 2D grating. 
As an example of the capabilities of the code and the method we used, we have also been able to produce 3D simulations of the collective scattering process. As the importance of polarization effects can be significant for 3D scattering, an extension of the present scalar model of light scattering to a vectorial model, simulating the 3D collective scattering from different atomic distribution and orientations of the pump, is in preparation.

\section{Acknowledgements}

This work was performed in the framework of the European Training Network ColOpt, which is funded by the European Union (EU) Horizon 2020 programme under the Marie  Sklodowska-Curie action, grant agreement 721465. R.A. thanks Funda\c{c}\~{a}o para a Ci\^{e}ncia e Tecnologia (FCT - Portugal) through the Ph.D. Grant PD/BD/105875/2014 (PD-F APPLAuSE).

\begin{widetext}
\appendix

\section{derivation of the motion equations}\label{appendix}

\subsection{Multimode collective recoil equations\label{multi:eqs}}

We consider the Hamiltonian of $N$ two-level atoms, with atomic frequency $\omega_a$ and dipole $d$, interacting with a laser field and the vacuum radiation modes:
\begin{equation}\label{Ham}
    H =\sum_{j=1}^N\frac{\mathbf{p}_j^2}{2M}+\hbar\sum_{j=1}^N\left[\frac{\Omega_0^*}{2}\sigma_{j}^- e^{i\Delta_0t-i\mathbf{k}_0\cdot\mathbf{r}_j}+\textrm{h.c.}\right]
    +\hbar\sum_{j=1}^N\sum_{\mathbf{k}}g_k\left[a_{\mathbf{k}}^\dagger \sigma_{j}^- e^{i\Delta_kt-i\mathbf{k}\cdot\mathbf{r}_j}
        +\sigma_{j}^+a_{\mathbf{k}}e^{-i\Delta_kt+i\mathbf{k}\cdot\mathbf{r}_j}\right],
\end{equation}
Here $\Omega_0=dE_0/\hbar$ is the Rabi frequency of the laser, with electric field $E_0$, wave vector $\mathbf{k}_0$ and frequency $\omega_0$, with detuning $\Delta_0=\omega_0-\omega_a$.  The quantum radiation modes in vacuum with wave
vectors $\mathbf{k}$ and frequency $\omega_k$ are described by the operators $a_\mathbf{k}$, with
$\Delta_k=\omega_k-\omega_a$, with coupling rate $g_k=d[\omega_k/(2\hbar\epsilon_0 V_{ph})]^{1/2}$, being $V_{ph}$ the quantization volume of the radiation field. We disregard polarization and short-range effects, using a scalar model for the radiation field. The internal dynamics of the two-level atoms are described by the operators
$\sigma_{j}^z=|e_j\rangle \langle e_j|- |g_j\rangle \langle g_j|$,~
$\sigma_{j}^+=|e_j\rangle\langle g_j|$ and
$\sigma_{j}^-=|g_j\rangle\langle e_j|$. Furthermore, we also consider the dynamics of the external degrees of freedom, where $\mathbf{r}_j$ and $\mathbf{p}_j$ are operators.
The Heisenberg equations are:
\begin{eqnarray}
\dot \mathbf{r}_j &=& \frac{\mathbf{p}_j}{M},\\
\dot \mathbf{p}_j &=& - \nabla_{\mathbf{r}_j}H\nonumber\\
&=&i\hbar \mathbf{k}_0\left[\frac{\Omega_0^*}{2}\sigma_{j}^- e^{i\Delta_0t-i\mathbf{k}_0\cdot\mathbf{r}_j}-\textrm{h.c.}\right]+i\hbar\sum_{\mathbf{k}}\mathbf{k}g_k\left[a_{\mathbf{k}}^\dagger\sigma_{j}^- e^{i\Delta_kt-i\mathbf{k}\cdot\mathbf{r}_j}-\textrm{h.c.}\right],\\
  \dot\sigma_{j}^- &=& \frac{i\Omega_0}{2}e^{-i\Delta_0 t+i \mathbf{k}_0\cdot \mathbf{r}_j}\hat\sigma_{j}^z +i\sum_{\mathbf{k}}g_k
  \sigma_{j}^z a_{\mathbf{k}}e^{-i\Delta_k t+i \mathbf{k}\cdot \mathbf{r}_j}\label{s1},\\
  \dot{\sigma}_{j}^z &=&  i\Omega_0^* e^{i\Delta_0 t-i \mathbf{k}_0\cdot \mathbf{r}_j}\sigma_{j}^- +2i\sum_{\mathbf{k}}g_k
  a_{\mathbf{k}}^\dagger\sigma_{j}^- e^{i\Delta_k t-i \mathbf{k}\cdot
  \mathbf{r}_j}+ \textrm{h.c.}\label{s3},\\
  \dot{a}_{\mathbf{k}} &=& -ig_k\sum_{j=1}^N \sigma_{j}^-e^{i\Delta_k t-i \mathbf{k}\cdot
  \mathbf{r}_j}.\label{ak}
\end{eqnarray}
Introducing $\sigma_j=\sigma_{j}^- e^{i\Delta_0 t}$ and neglecting the population of the excited state (assuming weak field and/or large detuning $\Delta_0$), so that $\sigma_{j}^z\approx -1$:
\begin{eqnarray}
\dot \mathbf{r}_j &=& \frac{\mathbf{p}_j}{M},\\
\dot \mathbf{p}_j &=&
i\hbar \mathbf{k}_0\left[\frac{\Omega_0^*}{2}\sigma_{j} e^{-i\mathbf{k}_0\cdot\mathbf{r}_j}-\textrm{h.c.}\right]+i\hbar\sum_{\mathbf{k}}\mathbf{k}g_k\left[a_{\mathbf{k}}^\dagger\sigma_{j} e^{i(\omega_k-\omega_0)t-i\mathbf{k}\cdot\mathbf{r}_j}-\textrm{h.c.}\right]\label{force:1},\\
  \dot\sigma_{j} &=& (i\Delta_0-\Gamma/2)\sigma_j-\frac{i\Omega_0}{2}e^{i \mathbf{k}_0\cdot \mathbf{r}_j}
  -i\sum_{\mathbf{k}}g_k
 a_{\mathbf{k}}e^{-i(\omega_k-\omega_0)t+i\mathbf{k}\cdot \mathbf{r}_j}\label{s1bis},\\
  \dot a_{\mathbf{k}} &=& -ig_ke^{i(\omega_k-\omega_0)t}\sum_{j=1}^N \sigma_{j}e^{-i\mathbf{k}\cdot
  \mathbf{r}_j}\label{akbis}.
\end{eqnarray}
where we added the spontaneous emission decay term $-(\Gamma/2)\sigma_j$, with $\Gamma=d^2k^3/2\pi\epsilon_0\hbar$ as the spontaneous decay rate. Assuming $\Gamma\gg\omega_{\mathrm{rec}}$, being $\omega_{\mathrm{rec}}=\hbar k^2/2M$ the recoil frequency, we can  adiabatically eliminate the internal degree of freedom, taking $\dot\sigma_j\approx 0$ in Eq.(\ref{s1bis}):
\begin{eqnarray}
\sigma_j &\approx & \frac{\Omega_0}{2(\Delta_0+i\Gamma/2)}e^{i \mathbf{k}_0\cdot \mathbf{r}_j}
  +\frac{1}{\Delta_0+i\Gamma/2}\sum_{\mathbf{k}}g_k
 a_{\mathbf{k}}e^{-i(\omega_k-\omega_0)t+i\mathbf{k}\cdot \mathbf{r}_j}\label{s1:adia}.
 \end{eqnarray}
The first term describes the dipole excitation induced by the driving field, whereas the second term is the excitation induced by the scattered field. By inserting it in Eq.(\ref{akbis}), the field equation, we obtain:
\begin{eqnarray}
  \dot a_{\mathbf{k}} &=& -i\frac{g_k\Omega_0}{2(\Delta_0+i\Gamma/2)}e^{i(\omega_k-\omega_0)t}\sum_{j=1}^Ne^{i(\mathbf{k}_0-\mathbf{k})\cdot\mathbf{r}_j}
  -i\frac{g_k}{\Delta_0+i\Gamma/2}\sum_{j=1}^N \sum_{\mathbf{k}'}g_{k'}
 a_{\mathbf{k}'}e^{i(\omega_k-\omega_{k'})t-i(\mathbf{k}-\mathbf{k}')\cdot \mathbf{r}_j}\label{akdot}.
\end{eqnarray}
The first term describes the single-scattering process, where the momentum  transfer to the atoms is from the incident field  to the vacuum field. The second term describes multiple scattering processes, where a photon is exchanged between the mode $\mathbf{k}$ and all the other modes $\mathbf{k}'$. We limit our analysis to single-scattering processing, neglecting the second term in Eq.(\ref{akdot}).
We also insert Eq.(\ref{s1:adia}) in the force equation (\ref{force:1}),
\begin{eqnarray}
\dot \mathbf{p}_j &=&
-\frac{i\hbar}{\Delta_0-i\Gamma/2}\left[\mathbf{k}_0\frac{\Omega_0}{2} e^{i\mathbf{k}_0\cdot\mathbf{r}_j}+\sum_{\mathbf{k}}\mathbf{k}g_k a_{\mathbf{k}}  e^{i\mathbf{k}\cdot\mathbf{r}_j-i(\omega_k-\omega_0)t}\right]
\left[
\frac{\Omega_0^*}{2}e^{-i \mathbf{k}_0\cdot \mathbf{r}_j}
  +\sum_{\mathbf{k}}g_k
 a_{\mathbf{k}}^\dagger e^{-i\mathbf{k}\cdot \mathbf{r}_j+i(\omega_k-\omega_0)t}\right]+\textrm{h.c.}\nonumber\\
\end{eqnarray}
The first and second terms in the first squared parenthesis describe the absorption of an incident photon with momentum $\hbar \mathbf{k}_0$ and a scattered photon with momentum $\hbar \mathbf{k}$, respectively. The second squared parenthesis is the response of the atom, i.e., the induced polarization of the atoms to the total radiation.  Explicitly, we write: 
\begin{eqnarray}
\dot \mathbf{p}_j
 &=&\left[\frac{\Gamma\Omega_0^2 }{4\Delta_0^2+\Gamma^2}\right]\hbar\mathbf{k}_0\nonumber\\
 &+&\frac{2i\Delta_0}{4\Delta_0^2+\Gamma^2}\sum_{\mathbf{k}}\hbar(\mathbf{k}_0-\mathbf{k})g_k\left[\Omega_0^*a_{\mathbf{k}}  e^{-i(\mathbf{k}_0-\mathbf{k})\cdot\mathbf{r}_j-i(\omega_k-\omega_0)t}-\textrm{h.c.}\right]\nonumber\\
 &+&\frac{\Gamma}{4\Delta_0^2+\Gamma^2}\sum_{\mathbf{k}}\hbar(\mathbf{k}_0+\mathbf{k})g_k\left[\Omega_0^*a_{\mathbf{k}}  e^{-i(\mathbf{k}_0-\mathbf{k})\cdot\mathbf{r}_j-i(\omega_k-\omega_0)t}+\textrm{h.c.}\right]\nonumber\\
 &+&\frac{1}{\Delta_0^2+\Gamma^2/4}\sum_{\mathbf{k}}\sum_{\mathbf{k}'}g_kg_{k'}\hat{a}_{\mathbf{k}}^\dagger\hat{a}_{\mathbf{k}'} e^{i(\mathbf{k}'-\mathbf{k})\cdot\mathbf{r}_j}\left[i\Delta_0
 \hbar(\mathbf{k}-\mathbf{k}')+(\Gamma/2)\hbar(\mathbf{k}+\mathbf{k}')\right].\label{force:2}
\end{eqnarray}
Notice that the first term is the radiation pressure exerted by the incident light (which is constant for a plane wave); the second and third terms describe the momentum transfer due to the exchange of photons between the incident and the scattered light. The last term is the contribution due to the exchange between two scattered vacuum photons of momentum $\hbar\mathbf{k}$ and $\hbar\mathbf{k}'$. Again, since we neglect multiple-scattering events, we drop the last term. Then, we assume $\Delta_0\gg\Gamma$, so that the first and the third terms of Eq.(\ref{force:2}) are negligibly small, thus, achieving:
\begin{eqnarray}
\dot \mathbf{p}_j
 &\approx &\frac{i}{2\Delta_0}\sum_{\mathbf{k}}\hbar(\mathbf{k}_0-\mathbf{k})g_k\left[\Omega_0^*a_{\mathbf{k}}  e^{-i(\mathbf{k}_0-\mathbf{k})\cdot\mathbf{r}_j-i(\omega_k-\omega_0)t}-\textrm{h.c.}\right].
\end{eqnarray}
The force on the atoms is the usual dipole (or gradient) force, where the momentum transfer is maximum for back-scattering emission (i.e., $\mathbf{k}=-\mathbf{k}_0$). In conclusion, the multi-mode equations describing the collective recoil are:
\begin{eqnarray}
\dot \mathbf{r}_j &=& \frac{\mathbf{p}_j}{M},\label{CARL:1}\\
\dot \mathbf{p}_j &=&
i\hbar g\sum_{\mathbf{k}}(\mathbf{k}_0-\mathbf{k})\left[A_{\mathbf{k}}  e^{-i(\mathbf{k}_0-\mathbf{k})\cdot\mathbf{r}_j}-
 A_{\mathbf{k}}^\dagger e^{i(\mathbf{k}_0-\mathbf{k})\cdot\mathbf{r}_j}\right],\label{CARL:2}\\
\dot{A}_{\mathbf{k}} &=& -ig\sum_{j=1}^Ne^{i(\mathbf{k}_0-\mathbf{k})\cdot\mathbf{r}_j}-i\delta_kA_\mathbf{k},\label{CARL:3}
\end{eqnarray}
where ${A}_{\mathbf{k}}=a_{\mathbf{k}}e^{-i\delta_kt}$, ~$\delta_k=\omega_k-\omega_0$ ~and~ $g=g_{k_0}(\Omega_0/2\Delta_0)$; we assumed $g_k\approx g_{k_0}$ and $\Omega_0$ real.

\subsection{Collective recoil equations in free space\label{coll:eqs}}

In free space the light is scattered in the 3D vacuum modes. Following ref.\cite{Moore}, we eliminate the scattered field by integrating  Eq.(\ref{CARL:3}) to obtain
\begin{equation}
A_{\mathbf{k}}(t) = A_{\mathbf{k}}(0)e^{-i(\omega_k-\omega_0)t}
-igN\int_0^t\rho_{\mathbf{k}_0-\mathbf{k}}(t-\tau)e^{-i(\omega_k-\omega_0)\tau}d\tau,\label{ak:int}
\end{equation}
with
\begin{equation}
\rho_{\mathbf{q}}(t)=\frac{1}{N}\sum_{j=1}^Ne^{i\mathbf{q}\cdot\mathbf{r}_j(t)}.\label{rho_q}
\end{equation}
The first term in Eq.(\ref{ak:int}) gives the free electromagnetic field,
i.e., vacuum fluctuations, and the second term is the
radiation field due to Rayleigh scattering.
If Eq.(\ref{ak:int}) is substituted into equation (\ref{CARL:2}) for $\mathbf{p}_j$, we obtain:
\begin{equation}
\dot \mathbf{p}_j =
\hbar g^2N\sum_{\mathbf{k}}(\mathbf{k}_0-\mathbf{k})\int_0^td\tau\left[\rho_{\mathbf{\mathbf{k}_0-\mathbf{k}}}(t-\tau) e^{-i(\mathbf{k}_0-\mathbf{k})\cdot\mathbf{r}_j}e^{-i(\omega_k-\omega_0)\tau}+\mathrm{h.c.}\right],\label{CARL:4:sr}
\end{equation}
where the first term of Eq.(\ref{ak:int}) has been neglected. Then, transforming the sum over $\mathbf{k}$ into an integral and using Eq.(\ref{rho_q}), we attain the coming expression:
\begin{equation}
\dot \mathbf{p}_j =
\hbar g^2\frac{V_{ph}}{8\pi^3}\sum_{m\neq j}\left[e^{-i\mathbf{k}_0\cdot(\mathbf{r}_j-\mathbf{r}_m)}\int_0^td\tau e^{i\omega_0\tau}\int d\mathbf{k}(\mathbf{k}_0-\mathbf{k}) e^{i\mathbf{k}\cdot(\mathbf{r}_j-\mathbf{r}_m)}e^{-ick\tau}+\mathrm{h.c.}\right],\label{CARL:5:sr}
\end{equation}
in which we used the Markov approximation so that $\mathbf{r}_j(t-\tau)\approx \mathbf{r}_j(t)$. The integral over $\mathbf{k}$, in the latter equation, can be manipulated as follow:
\begin{eqnarray}
\int d\mathbf{k}(\mathbf{k}_0-\mathbf{k}) e^{i\mathbf{k}\cdot(\mathbf{r}_j-\mathbf{r}_m)}e^{-ick\tau}&=&4\pi \mathbf{k}_0\int_0^\infty dk k^2\frac{\sin(kr_{jm})}{kr_{jm}}e^{-ick\tau}\nonumber\\
&+&4i\pi\hat\mathbf{r}_{jm}\int_0^\infty dk k^3
\left[\frac{\cos(kr_{jm})}{kr_{jm}}-\frac{\sin(kr_{jm})}{(kr_{jm})^2}\right]e^{-ick\tau},
\end{eqnarray}
being $\mathbf{r}_{jm}=\mathbf{r}_{j}-\mathbf{r}_{m}$, ~$r_{jm}=|\mathbf{r}_{jm}|$ and $\hat\mathbf{r}_{jm}=\mathbf{r}_{jm}/r_{jm}$. Since $k\approx k_0$, we can replace  $k$  by $k_0$ in the integrals; we also extend the lower integration limit to $-\infty$, reaching the next expression:
\begin{eqnarray}
\int d\mathbf{k}(\mathbf{k}_0-\mathbf{k}) e^{i\mathbf{k}\cdot(\mathbf{r}_j-\mathbf{r}_m)}e^{-ick\tau}&\approx &4\pi k_0^3\frac{\hat\mathbf{z}}{k_0r_{jm}}\int_{-\infty}^\infty dk \sin(kr_{jm})e^{-ick\tau}\nonumber\\
&+&4i\pi k_0^3\frac{\hat\mathbf{r}_{jm}}{k_0r_{jm}}\int_{-\infty}^\infty dk
\left[\cos(kr_{jm})-\frac{\sin(kr_{jm})}{k_0r_{jm}}\right]e^{-ick\tau}\nonumber\\
&=&\frac{4\pi^2 k_0^3}{c}\left\{\frac{\hat\mathbf{z}}{ik_0r_{jm}}
\left[\delta(\tau-r_{jm}/c)-\delta(\tau+r_{jm}/c)\right]\right.\nonumber\\
&-&\left.\frac{\hat\mathbf{r}_{jm}}{ik_0r_{jm}}
\left[\delta(\tau-r_{jm}/c)+\delta(\tau+r_{jm}/c)\right]\right.\nonumber\\
&-&\left.\frac{\hat\mathbf{r}_{jm}}{(k_0r_{jm})^2}
\left[\delta(\tau-r_{jm}/c)-\delta(\tau+r_{jm}/c)\right]\right\},\label{integral}
\end{eqnarray}
where we assumed $\mathbf{k}_0=k_0\hat\mathbf{z}$ and used the following two integrals:
\begin{eqnarray}
\int_{-\infty}^\infty dk \sin(kR)e^{-ick\tau}&=&\frac{\pi}{ic}\left[\delta(\tau-R/c)-\delta(\tau+R/c)\right],\nonumber\\
\int_{-\infty}^\infty dk \cos(kR)e^{-ick\tau}&=&\frac{\pi}{c}\left[\delta(\tau-R/c)+\delta(\tau+R/c)\right].\label{dirac}
\end{eqnarray}
By inserting Eq.(\ref{integral}) into Eq.(\ref{CARL:5:sr}), together with the definitions of $g$ and $\Gamma$, we are able to derive the final expression for the force:
\begin{eqnarray}
\dot \mathbf{p}_j &=&
\frac{\Gamma}{2}\hbar k_0\left(\frac{\Omega_0}{2\Delta_0}\right)^2\sum_{m\neq j}\left[e^{ik_0(r_{jm}-z_{jm})}
\left[
\frac{(\hat\mathbf{z}-\hat\mathbf{r}_{jm})}{ik_0r_{jm}}-\frac{\hat\mathbf{r}_{jm}}{(k_0r_{jm})^2}
\right]
+\mathrm{h.c.}\right]\nonumber\\
&=&\Gamma\hbar k_0\left(\frac{\Omega_0}{2\Delta_0}\right)^2\sum_{m\neq j}\left\{
(\hat\mathbf{z}-\hat\mathbf{r}_{jm})
\frac{\sin[k_0(r_{jm}-z_{jm})]}{k_0r_{jm}}-\hat\mathbf{r}_{jm}\frac{\cos[k_0(r_{jm}-z_{jm})]}{(k_0r_{jm})^2}
\right\}.
\label{CARL:3:sr}
\end{eqnarray}

\subsection{Radiation field}

The scattered radiation field amplitude is
\begin{equation}
 E_s(\mathbf{r},t)=i\frac{V_{ph}}{(2\pi)^3}e^{i(\mathbf{k}_0\cdot\mathbf{r}-\omega_0 t)}\int_{\Delta\mathbf{k}} d\mathbf{k} {\cal E}_k A_\mathbf{k}(t)e^{i(\mathbf{k}-\mathbf{k}_0)\cdot\mathbf{r}},
 \label{Es}
 \end{equation}
being ${\cal E}_k=(\hbar\omega_k/2\epsilon_0 V_{ph})^{1/2}$ the 'single-photon' electric field.
Using eq.(\ref{ak:int}), neglecting the fluctuation term and transforming the sum over $\mathbf{k}$ into an integral, as done before, we obtain
\begin{eqnarray}
 E_s(\mathbf{r},t)
&=&\frac{V_{ph}}{(2\pi)^3}ge^{-i\omega_0 t}\sum_{j=1}^N\int_0^t d\tau e^{i\omega_0\tau}e^{i\mathbf{k}_0\cdot\mathbf{r}_j(t-\tau)}\int_0^\infty dk k{\cal E}_{k}\frac{\sin(k|\mathbf{r}_j(t-\tau)-\mathbf{r}|)}{|\mathbf{r}_j(t-\tau)-\mathbf{r}|}
e^{-ick\tau}.
 \end{eqnarray} 
The scattered intensity will be centered about the incidence laser frequency $\omega_0$. The quantity $ck$ varies little around $k=\omega_0/c$ for which the time
integral in $\tau$ is not negligible. We can therefore replace $k$ by $\omega_0/c$ and extend
the lower limit in the $k$ integration by $-\infty$:
\begin{eqnarray}
 E_s(\mathbf{r},t)&=&\frac{V_{ph}}{2\pi^2}gk_0{\cal E}_{k_0}e^{-i\omega_0 t}\sum_{j=1}^N\int_0^t d\tau e^{i\omega_0\tau+i\mathbf{k}_0\cdot\mathbf{r}_j(t-\tau)}\int_{-\infty}^\infty dk \frac{\sin(k|\mathbf{r}_j-\mathbf{r}|)}{|\mathbf{r}_j-\mathbf{r}|}
e^{-ick\tau}.
 \end{eqnarray}
By using Eq.(\ref{dirac}) we obtain:
\begin{eqnarray}
 E_s(\mathbf{r},t)&=&\frac{dk_0^3}{4\pi\epsilon_0}\frac{\Omega_0}{2\Delta_0}\sum_{j=1}^N\frac{e^{ik_0R_j}}{ik_0R_j}e^{i(\mathbf{k}_0\cdot\mathbf{r}_j-\omega_0t)}\Theta(t>R_j/c),
 \end{eqnarray}
where  $R_j=|\mathbf{r}_j-\mathbf{r}|$ and $\mathbf{r}_j$ is evaluated at the retarded time $t-R_j/c$.
Assuming $\mathbf{r}\gg\mathbf{r}_j$, we can write $R_j\approx r-i\hat\mathbf{r}\cdot\mathbf{r}_j$ with $\hat \mathbf{r}=\mathbf{r}/r$, and
\begin{eqnarray}
 E_s(\mathbf{k},t)&\approx &\frac{dk_0^2}{4\pi\epsilon_0}\frac{\Omega_0}{2\Delta_0}\frac{e^{i(k_0r-\omega_0t)}}{ir}\sum_{j=1}^N e^{i(\mathbf{k}_0-\mathbf{k})\cdot\mathbf{r}_j},
 \end{eqnarray}
where $\mathbf{k}=k_0\hat\mathbf{r}$.
We have obtained the expression of the Rayleigh scattering field in the far-field limit, i.e., a spherical wave proportional to the factor form, depending on the geometrical configuration of the scattering particles. For small clouds we can neglect the retarded time $R_j/c$.
In conclusion, the scattered intensity spatial distribution in the far-field limit is
\begin{eqnarray}
 I_s(\mathbf{k})&=&I_1N^2 |M(\mathbf{k},t)|^2,
 \end{eqnarray}
 in which $I_1=(\hbar\omega_0\Gamma/8\pi r^2)(\Omega_0/2\Delta_0)^2$ is the single-atom Rayleigh scattering intensity and
\begin{eqnarray}
M(\mathbf{k},t)&=&\frac{1}{N}\sum_{j=1}^N e^{i(\mathbf{k}_0-\mathbf{k})\cdot\mathbf{r}_j(t)}
 \end{eqnarray}
is the 'optical magnetization', or 'bunching factor'.
\end{widetext}

\end{document}